\documentclass[aps,prd,twocolumn,groupedaddress,showpacs,preprintnumbers,draft]
{revtex4}
\usepackage{epsf}
\usepackage{bm}
\usepackage{amsfonts}

\newcommand{\beqa}{\begin{eqnarray}}
\newcommand{\eeqa}{\end{eqnarray}}
\newcommand{\beq}{\begin{equation}}
\newcommand{\eeq}{\end{equation}}
\newcommand{\bal}{\begin{align}}
\newcommand{\eal}{\end{align}}

\def\bear{\begin{array}}
\def\enar{\end{array}}

\def\mphi{m_{\phi}}
\def\mchi{m_{\chi}}
\def\mi{m_I}
\def\h2{\frac{\mi^2}{M_{p}}}
\def\Mp{M_{p}}

\def\bear{\begin{array}}
\def\enar{\end{array}}

\begin{document}

\preprint{\vbox{ \hbox{} \hbox{hep-ph/0507xxx} \hbox{July 2005} }}


\title{Non-Perturbative Instabilities as a Solution of the 
Cosmological Moduli Problem}

\author{Natalia Shuhmaher \email[Email:]{shuhmaher@hep.physics.mcgill.ca}
and Robert Brandenberger \email[Email:]{rhb@hep.physics.mcgill.ca}}
\affiliation{Department of Physics, McGill University, Montr\'eal, QC,
H3A 2T8, CANADA}



\begin{abstract} 

It is widely accepted that moduli in the mass range $10$eV -
$10^4$GeV which start to oscillate with an amplitude of the order
of the Planck scale either jeopardize successful predictions of
nucleosynthesis or overclose the Universe. It is shown that the
moduli problem can be relaxed by making use of parametric
resonance. A new non-perturbative decay channel for moduli
oscillations is discussed. This channel becomes effective when the
oscillating field results in a net negative mass term for the
decay products. This scenario allows for the decay of the moduli
much before nucleosynthesis and, therefore, leads to a complete
solution of the cosmological moduli problem.

\end{abstract}


\maketitle

\section{Introduction}

In many theories beyond the Standard Model of particle physics, in
particular in supergravity and string theories, there are many
scalar and fermionic fields with masses smaller or equal to the
electroweak scale and gravitational strength couplings to ordinary
matter. Such fields are called {\it moduli fields} and behave as
non-relativistic matter at late time. Since they decay late because
of their weak interactions, they lead to the so-called
cosmological moduli problem
\cite{Coughlan,Ellis2,Carlos,Banks:1993en}.
Below, independently of their origin, fields having only Planck
scale couplings and weak scale mass are denoted collectively as
moduli fields.

In the case of moduli in supersymmetric theories, the mass
$m_{\phi}$ of these fields $\phi$ is generated during
supersymmetry breaking. A wide variety of these scenarios predict
masses of the moduli in the dangerous range, $m_{\phi}
\sim 10eV - 10^4 GeV$. According to our ideas of early Universe
cosmology, we expect moduli to be produced in great abundance in
the early universe. In the context of Big Bang cosmology, both
scalar and fermionic moduli particles will be part of the initial
thermal bath of particles of the very early universe. Even
assuming that the moduli particles are not part of the initial
thermal bath (for example in the context of inflationary
cosmology) it is hard to avoid the presence of excited moduli
fields at late times. For example, in the case of scalar moduli,
since the moduli are massless before supersymmetry breaking, there
is no reason that the moduli field values before supersymmetry
breaking coincide with the values which turn into the minima of
the potential after supersymmetry breaking
\cite{Goncharov}. The offset will lead to moduli fields which
oscillate about their potential minima. An offset of a scalar
modulus can also be produced by quantum fluctuations in the early
phases of inflation, as follows from the computation of the
coincident point two point function of a low-mass scalar field
during inflation
\cite{Linde,FordVil}. The excessive production of moduli is
predicted during the preheating stage of inflationary cosmology in
a wide variety of models
\cite{Giudice2}. A further source of moduli particles is
gravitational particle production between the end of inflation and
the time of nucleosynthesis
\cite{Giudice1,Felder2}.

Due to their weak interactions, the decay of the moduli fields is
slow. Widely used estimates based on dimensional analysis for the
perturbative decay rate $\Gamma$ give
\beq \label{decay}
\Gamma \, ~\sim \, {{\mphi^3} \over {\Mp^2}} \,  .
\eeq
The presence of excited moduli fields at late times is dangerous
since the presence of the extra moduli field energy during the
time of nucleosynthesis could spoil the success of the standard
Big Bang nucleosynthesis scenario
\cite{Ellis1}. This danger is acute in particular for the heavier
moduli fields. Both lighter moduli fields and heavier fields which
do not decay before the time of equal matter and radiation threaten to
overclose the Universe at that time (and thus also today if they
do not decay between the time of equal matter and radiation and the present
time).

To be more specific, it is the fact that the interactions of
moduli fields come from non-renormalizable terms in the
supersymmetry (SUSY) potential that leads to the moduli problem.
For example, the following term in the K\"ahler potential
\beq \label{Kahler}
\delta K = {1 \over \Mp^2} \phi_I^\dagger \phi_I \phi^\dagger \phi \, ,
\eeq
results in a contribution to the square mass for $\phi$ of the
form $\rho/\Mp^2$~\cite{Dine:1995uk}. By $\phi$ we denote a
canonically normalized modulus field (with bare mass $\mphi$), by
$\Mp$ - the Planck mass, by $\phi_I$ - a field which dominates the
energy density of the universe, the inflaton, and by $\rho$ - the
energy density contained in the inflaton. Since $\rho = 3 H^2
\Mp^2$, the effective mass $m_{eff}$ becomes
\beq
m_{eff}^2 = c H^2 + \mphi^2 \, ,
\eeq
where $c$ is a constant. If $c \gg 1$, then inflation drives the
moduli to the minima of
their high temperature effective potential. However, typically the
high temperature minima are offset from the zero temperature
minima of the moduli potentials by a value which has Planck order
of magnitude. If $c \ll 1$ then $ H > m_{eff}$, and then quantum
fluctuations during inflation will also excite the field $\phi$ to
a value of the order of the Planck mass.  During reheating the
energy density of the inflaton stops dominating the Universe and
the effective mass of $\phi$ relaxes to $m_\phi$. After reheating,
the Hubble constant decreases in the radiation dominated phase.
Once $H \sim m_\phi$ the condition for slow rolling of $\phi$
($V''(\phi)< H^2$) is no longer satisfied, and at that point the
field $\phi$ starts to oscillate around its low temperature
minimum which we take to be zero. The energy density of $\phi$
decreases like that of non-relativistic particles (i.e.
proportional to $a(t)^{-3}$, where $a(t)$ is the cosmological
scale factor), whereas the energy density of the radiation
dominated universe falls as $a(t)^{-4}$. Thus, the modulus field
$\phi$ may come to dominate the energy density of the universe, or
at least contribute too much during the period of nucleosynthesis,
unless the field decays early. The problem is that the modulus
fields are coupled only gravitationally to themselves and other
particles and thus have very small perturbative decay rates.

There has been some previous work to try to mitigate the
cosmological moduli problem. Entropy production at late times will
dilute the moduli density. For example, weak scale inflation
\cite{Randall:1994fr} or thermal inflation
\cite{Stewart} could sufficiently dilute unwanted moduli
(see e.g. \cite{Asaka:1999xd} for a detailed study of the
potential of thermal inflation to solve the moduli problem for
masses in the above-mentioned dangerous range predicted in models
with hidden sector and gauge-mediated supersymmetry breaking). For
certain ranges of parameters of a gauge-mediated supersymmetry
breaking model, the oscillations of the modulus field itself might
sufficiently dilute the string moduli density
\cite{Kawasaki1}. The decay of an unstable domain wall network
\cite{Kawasaki2} is another
way to generate entropy and dilute the moduli density at late
times. A common danger of these approaches is that the baryon
density might also be diluted to an unacceptably low value.
Another approach is to invoke effects which give the moduli fields
a contribution to the square mass of the order of $~H^2$ which
would allow them to roll down their potential during inflation
\cite{Dine84,Coughlan84} and prevent them from acquiring a large
expectation value during inflation (see also \cite{Linde96}).
However, this solution does not work if the low temperature
minimum of the moduli potential does not coincide \cite{Goncharov}
with its high-temperature ground state (new symmetries which could
force the two states to be the same were analyzed in
\cite{Dine:1995uk}). A recent proposal to solve the moduli problem
is moduli trapping at enhanced symmetry points
\cite{Kofman,Watson}. In terms of the use of parametric resonance
instabilities, our work has similarities with that of
\cite{Kofman,Watson}. However, in contrast to these works, in
our study the focus is on the traditional moduli problem as
formulated in \cite{Coughlan,Ellis2,Carlos,Banks:1993en}.

In this paper, we propose a way to solve the cosmological problems
of scalar moduli fields which requires no external mechanism for
the dilution of moduli. Instead, it makes use of non-perturbative
decay channels. Non-perturbative decays have been shown to
completely change the scenario of reheating in inflationary
cosmology. In particular, the decay of the inflaton field by a
parametric resonance instability has been shown to be very
important~\cite{TB,KLS1,STB,KLS2}. In certain models, a tachyonic
instability renders the inflaton decay even more
efficient~\cite{Felder}.

We consider a toy model which potentially gives rise to a
cosmological moduli problem. In the framework of this model we
investigate non-perturbative decay channels (decays into 
particles whose lifetime is much shorter than that of the
moduli fields). We study the decay of
the oscillating modulus field via parametric resonance, and
propose a new tachyonic decay channel. We quantify the conditions
on the parameters of the model for which the decay channels are
effective. Note that the decay channels work for an initial field
amplitude up to the order of the Planck scale. No external
mechanism for diluting the moduli density is required.

\section{The Model}

If the moduli problem arises as a consequence of supersymmetry
breaking, the moduli potential takes the form (see e.g.
\cite{Randall:1994fr})
\beq
{\cal V}(\Phi) \, = \, m^2_{3/2} \Mp^2 {\cal G} (|\Phi|/\Mp) \, ,
\label{eq:potential}
\eeq
where $m_{3/2}$ is the gravitino mass and ${\cal G}$ is some
function. We will assume that the modulus field couples to some
matter field $\chi$ which we treat as a scalar field (following
the analyses in the study of fermionic preheating \cite{Fermionic}
it could also be taken to be a fermionic field). Making use of
(\ref{eq:potential}), and of the fact that the dimension five and
six operators are suppressed by $\Mp$ and by $\Mp^2$,
respectively, we can construct a typical potential for the modulus
$\phi$ and the matter scalar field $\chi$
\begin{eqnarray}
V \, = \,&& \frac{1}{2} \mphi^2 \phi^2 + \frac{1}{2} \mchi^2 \chi^2
+ {1 \over 2}\h2 \phi \chi^2 \\
&& + \lambda_1 \phi^2 \chi^2 + \lambda_2 \chi^4 + 
\lambda_3 \phi^4 \, , \nonumber
\end{eqnarray}
where the coupling constants $\lambda_1 ,\ \lambda_2 , \
\lambda_3$ are small enough such that for $|\phi|, \ |\chi| < \Mp$
the low energy effective potential for $\phi$ takes form
\beq
V_l \, = \, \frac{1}{2} \mphi^2 \phi^2 + \frac{1}{2} \mchi^2
\chi^2 + {1 \over 2}\h2 \phi \chi^2 \, .
\eeq
Thus, the toy model we consider is described by the following
Lagrangian:
\begin{eqnarray} \label{lagr}
L \, = \, && \partial^\mu\phi\partial_\mu\phi +
\partial^\mu\chi\partial_\mu\chi \\
&& - \frac{1}{2} \mphi^2 \phi^2 -
\frac{1}{2} \mchi^2 \chi^2 - {1 \over 2}\h2 \phi \chi^2 \, . \nonumber
\end{eqnarray}
In the above, the mass $\mi$ sets the scale of the
interaction between $\phi$ and $\chi$. It is reasonable to assume
that this scale is much less than $M_p$.

In analogy to the situation encountered in the study of the decay
of the inflaton, if we want to study the decay of the modulus
field, it is crucial to focus on the equation of motion for matter
fields $\chi$ which the modulus field couples to. In the presence
of the oscillating modulus field, this equation
(in an expanding space-time) is
\begin{eqnarray}
\ddot{ \chi}_k &+& 3H \dot{\chi}_k + \left( {{k^2} \over {a^2}} +\mchi^2 +
\h2\Phi(t) \sin(\mphi t) \right) \chi_k \, \nonumber \\
&=& \, 0 \, ,
\end{eqnarray}
where $\Phi(t)$ is the amplitude of $\phi$. The amplitude $\Phi$
decreases as a consequence of the expansion of space. In the
above, $k$ denotes the comoving wavenumber and $a(t)$ is the
cosmological scale factor.

In a first step, we will put the above equation into the form of
the well-known Mathieu equation. To absorb the expansion of space,
we define a rescaled field via
\beq
\eta_k \, = \, a^{3/2} \chi_k \, .
\eeq
Then, the equation of motion for $\eta_k$ becomes:
\beq \label{etaeq1}
\ddot{ \eta}_k +  \left( {{k^2} \over {a^2}} +\mchi^2 +
\h2\Phi(t) \sin(\mphi t) - \Delta  \right) \eta_k \, = \, 0 \, ,
\eeq
where
\beq
\Delta \, = \, {3 \over 4} H^2 + {3 \over 2} {{\ddot a} \over a} \,
= \, {3 \over 2}(1/2 +(p-1)/p)H^2 \, ,
\eeq
where for the second equality we have assumed that $a(t) \propto
t^p$.

It is convenient to introduce a dimensionless time variable via
\beq
z \, = {1 \over 2} \mphi t + {{\pi} \over 4} \, .
\eeq
The differentiation with respect to $z$ will be denoted by a
prime. In this case, the above equation (\ref{etaeq1}) takes the
form
\beq \label{etaeq2}
\eta''_k + (A_k - 2q \cos 2z)\eta_k \, = \, 0
\, , \eeq
where
\begin{eqnarray}
A_k \, &=& \, 4 {{k^2} \over {\mphi^2 a^2}} + 4 {{\mchi^2} \over {\mphi^2}}
- 4 \mphi^{-2} \Delta \label{Aeq}\\
q \, &=& 2 {{\mi^2 \Phi}  \over {\mphi^2 \Mp}} \, . \label{qeq}
\end{eqnarray}

If it is justified to neglect the expansion of space compared to
the rate of the processes which will be discussed in the following,
then $\Delta$ vanishes, and the equation (\ref{etaeq2}) takes the
form of the Mathieu equation. Note that $k/a$ is the physical wavenumber.

%
\section{Parametric Resonance Instability}

The modulus field $\phi$ is frozen until the Hubble parameter $H$
drops to a value comparable to $\mphi$. Then, $\phi$ begins to
oscillate about $\phi = 0$ with a frequency $\mphi$, its amplitude
$\Phi(t)$ being damped by the expansion of the spatial background
and by the energy loss of $\phi$ to other fields. The second
effect is a back-reaction effect which we will neglect.
The condition required that the oscillation of $\phi$ begins before
the time of nucleosynthesis is that the Hubble damping term in the equation
of motion for $\phi$ becomes smaller than the force term $V^{'}(\phi)$
driving the oscillations. It yields
\beq
H(T_{NS}) \, < \, \mphi \, ,
\eeq
(where $T_{NS}$ is the temperature at which nucleosynthesis takes
place) a condition which is satisfied for all masses in the
dangerous range. Once the modulus field starts to oscillate,
resonant excitation of all fields coupled to $\phi$ is possible,
in particular the excitation of $\eta$.

The first instability we will study is the parametric resonance
instability, first applied to the decay of the inflaton field in
\cite{TB}.  There are two types of resonance \cite{KLS1}, {\it
broad parametric resonance} and {\it narrow parametric resonance}.
The condition for broad resonance is $q > 1$. where $q$ is the
parameter appearing in equation (\ref{etaeq2}). This condition is
satisfied for large values of the amplitude $\Phi$, namely for
\beq
\Phi \, > \, \Phi_b \equiv M_p \bigl( {{\mphi} \over {\mi}} \bigr)^2 \, .
\eeq
Evaluating this condition at the time when perturbative decay
of the modulus field sets in, i.e. when $\Gamma = H$,
and using for formula for $\Gamma$ applicable to our toy
model (given below in (\ref{decay2})), we find that unless
\beq
\mphi \, \ll \, \mi \bigl( {{\mi} \over {10M_p}} \bigr)^{1/2} \, ,
\eeq
broad parametric resonance can relax but not solve the moduli
problem without an additional decay channel being present. Both
narrow resonance (discussed below) and the tachyonic decay
discussed in the following section can provide the additional
channel.

For smaller values of $\Phi$, we are in the domain of narrow
parametric resonance. In this phase, the growth of $\eta_k$ is
known \cite{Landau,Arnold}:
\beq \label{growth}
\eta_k \, \sim \, e^{q z} \, \sim \, e^{q \mphi t / 2} \, .
\eeq
Only modes in narrow resonance bands are amplified, and the
first such band is centered at a value of $k = k_m$
given by \cite{Landau,Arnold}
\beq
A_k \, = \, 1 \, , \,\,\,
k_m \, = \, {\mphi \over 2}(1-\frac{4 \mchi^2}{\mphi^2})^{1/2}
\eeq
from which it follows that the band does not exist unless
$m_{\chi} < \mphi/2$. Other resonance bands occur for larger
values of $A_k$ but are of higher order in perturbation theory and
hence have a negligible effect.

The first condition for resonance to be effective is that the
typical time scale for the growth of $\eta_k$ is shorter than the
Hubble time, i.e.
\beq
q \mphi \, > \, H \, . \label{eq:light_condition}
\eeq
In addition, one should take into account the change of the
momentum as a result of the background expansion~\cite{KLS2}. In a
time interval $\delta t$, assuming $m_{\chi} < \mphi/4$, the
change in the physical momentum
$p = k/a(t)$ corresponding to the middle of the lowest resonance band
is
\beq
\Delta p  \, = \,  p H \Delta t \, \simeq \, \frac{\mphi}{2} H \Delta t  \, .
\eeq

The width of the resonance band is
\beq
\Delta p \, = \, {q \mphi \over 2} \, .
\eeq
Thus, $p$ remains in the resonance band during a time interval
\beq
\Delta t \, \simeq \, q H^{-1} \, .
\eeq
To justify neglecting the expansion of space, we must require that
the exponent in the growth factor (\ref{growth}) is at least $1$
during this time interval. This leads to a more severe constraint on
$q$:
\beq
q^2 \mphi \,  > \, H \, . \label{eq:severe_condition}
\eeq
Inserting the value of $q$ from (\ref{qeq}), we find that narrow
resonance is efficient provided
\beq
\Phi(t) \, > \, \Phi_c(t)
\equiv {{\sqrt{H} M_p \mphi^{3/2}} \over {\mi^2}} \, .
\eeq

Since $\sqrt{H}$ decreases as $t^{-1/2}$ whereas in a
radiation-dominated phase $\Phi(t)$ decreases only as $t^{-3/4}$,
the narrow resonance decay channel eventually shuts off. In order
that the moduli field does not dominate the energy density at the
time of the shutoff at temperature $T$, the following condition
needs to be satisfied:
\beq \label{phic}
\mphi^2 \Phi_c^2 \, \ll T^4 \, ,
\eeq
which, inserting the expression (\ref{phic}) for $\Phi_c$, turns
into the condition
\beq \label{bound}
\mphi \, \ll \, \mi \bigl( {{T^2} \over {\mi M_p}} \bigr)^{1/5} \, .
\eeq

For narrow parametric resonance to solve the
cosmological moduli problem, one needs to check that at the time of
moduli decay (which occurs when $\Gamma \sim H$),
the condition~(\ref{bound}) still holds. The decay rate
of $\phi$ for our Lagrangian (\ref{lagr}) is given by
\beq \label{decay2}
\Gamma \, = \, {{\mi^4} \over {32 \pi M_p^2 \mphi}} \, .
\eeq
Thus, the solution of the cosmological moduli problem requires:
\beq \label{mphi_range}
\mphi \, \ll \, \mi \bigl( {{\mi} \over {10 M_p}} \bigr)^{1/3} \, .
\eeq
Inserting the temperature of nucleosynthesis to get lower bound
from (\ref{decay2}) on the potentially dengerous mass range,
and using the value
$ \mi = 10^{2} $ GeV, the problem is solved for values of $\mphi$
satisfying
\beq
10^{-7}GeV < \mphi < 10^{-4} GeV\, .
\eeq
In the context of our model (for $\mi = 10^2$GeV), moduli with masses
smaller than $10^{-7}$GeV decay before time of nucleosynthesis
and thus do not cause the problem. Note that this scaling of the
decay rate with the mass of the decaying particle is going against
the intuition that lighter moduli should decay later than heavier one.
This curious aspect of our toy model decreases the potentially dangerous
mass range, and this realization might be useful in some concrete
models suffering from a moduli problem.

It appears at this point of our study that the period of narrow parametric
resonance has the potential of solving the moduli problem
for values of $\mphi$ and $\mi$ which satisfy the relation given by
(\ref{mphi_range}). One issue which we have not taken into account
is the fact that late moduli decay may provide a large source
of non-thermal photons which could distort the black-body
nature of the CMB. The constraints resulting from this effect
must be studied in any concrete model with late-decaying moduli
fields.

The previous analysis has missed a second important condition
for the efficiency of narrow parametric resonance. It is not
sufficient that the modes $\eta_k$ increase with a rate faster
than $H$. Since  the resonance occurs only in narrow bands
\cite{Landau,Arnold}, it is important to check that the rate of
energy increase integrated over all modes of $\eta$ be larger than
the decrease in the energy density of $\phi$ taking into account
the expansion of space alone. Otherwise, the energy stored
in the moduli field would still scale as matter in spite of the
exponential increase in the occupation number of certain field
modes. This condition reads
\beq \label{cond2}
{\dot{\rho_{\eta}}} \, > \, H \rho_{\phi} \, .
\eeq
The rate of increase ${\dot{\rho_{\eta}}}$ in the energy density of
$\eta$ can be estimated by considering the increase in the
amplitude of all modes of $\eta$ in the lowest instability band.
This band is located at $k \sim \mphi$ and its width is given by
$q \mphi$. Since the rate of increase (from (\ref{growth})) is
$q \mphi$ and since the initial mode (vacuum) energy is about $k$,
we obtain
\beq
{\dot{\rho_{\eta}}} \, \sim \, \mphi^5 q^2 \, .
\eeq
Hence, the condition (\ref{cond2}) for efficiency of the resonance
process becomes
\beq \label{cond5}
\mphi \, < \, \mi {{\mi^2} \over {T^2}} {{\mi} \over {M_p}} \, ,
\eeq
which is to be evaluated at the temperatures when the presence of
moduli fields are dangerous for cosmology. Using, as before, the
value $\mi = 10^2$GeV, and evaluating the above condition at the
temperature of nucleosynthesis, we find that the condition is
satisfied as long as the mass $\mphi$ is smaller than about
$10^{-5}$GeV. The condition (\ref{cond5}) becomes increasingly
well satisfied at lower temperatures, and is no longer a concern
at the time of recombination. Note that the condition (\ref{cond2})
for the efficiency of narrow resonance is a very conservative one.
As long as the first condition (\ref{eq:severe_condition}) is satisfied,
the mode amplitude will grow, and hence the vacuum mode amplitude
used in the above estimate should be replaced by the excited
amplitude, thus relaxing the constraint by an exponential factor.

The problem, however, is that the condition (\ref{cond5})
conflicts up to coefficients of order unity with the condition
that the perturbative decay rate is negligible. Thus, in our toy
model, and using the very conservative form of our conditions
for the effectiveness of the resonance process, 
the narrow resonance decay channel is only effective near
the time when the perturbative decay is also become important.
This result, however, is a consequence of the particular scaling
of $\Gamma$ with $\mphi$, $\Gamma \propto \mphi^{-1}$. In models
where moduli decay to fermions, $\Gamma \propto \mphi$, and,
therefore, we expect those models do not suffer from this specific
problem.

The above discussions neglected the expansion of the universe.
Taking into account this expansion changes the Mathieu equation
into a more general equation of Floquet type, and leads to a
stochastic nature of the resonance process \cite{KLS2}. However,
the property that the number of particles is growing exponentially
at a rate given by (\ref{growth}) is preserved.

%
\section{Tachyonic Decay of the Oscillating Modulus Field}

In the case of the decay of the inflaton field at the end of the
period of inflation, it is known \cite{Felder} that for certain
models there is a tachyonic instability channel which is more
efficient than parametric resonance. In this section, we will
study a similar process for moduli decay.

Let us return to the basic equation (\ref{etaeq2}), with the
values of the parameters $A_k$ and $q$ given by (\ref{Aeq}) and
(\ref{qeq}), respectively. We immediately see that for large
values of $\Phi$, the effective $m^2$ term in the equation will be
negative for part of the oscillation period of $\phi$. This
tachyonic instability occurs provided
\beq \label{cond3}
{{\mi^2 \Phi} \over {\mphi^2 M_p}} \, > \, {{m_{\chi}^2} \over
{\mphi^2}} \, .
\eeq
The minimal value for which the tachyonic decay channel is open is
given by setting the two sides in (\ref{cond3}) equal and will be
denoted by $\Phi_m$.

The condition under which the tachyonic decay channel can solve
the moduli problem is then given by the condition
\beq \label{cond4}
\mphi^2 \Phi_m^2 \, \ll \, T^4 \, ,
\eeq
where $T$ is the temperature corresponding to the period one is
interesting in. Evaluating (\ref{cond4}) at the time of
nucleosynthesis, we obtain
\beq \label{tachbound}
\mphi \, \ll \, T_{NS} {{T_{NS}} \over {M_p}} {{\mi^2} \over {m_{\chi}^2}} \, .
\eeq
The upper mass bound on $\mphi$ for which the above tachyonic
decay is effective thus depends sensitively on the ratio of $\mi$
and $m_{\chi}$. Unless the latter mass is much smaller than the
former, the tachyonic decay channel cannot reduce the amplitude of
moduli oscillations to a level consistent with the observational
constraints.

The range of values of $\mphi$ for which the two decay channels -
narrow parametric resonance (neglecting for a moment the issue
that in our model it starts to be efficient together with the
perturbative decay) and tachyonic decay - are open depends on the
values of the masses. While if $m_{\chi} > \mphi$, the only
allowed channel is the tachyonic one, the narrow parametric
resonance works for a wider range of masses $\mphi$ when $m_{\chi}
\sim \mphi$. The later can be seen by inserting $m_{\chi} = \mphi$
into (\ref{tachbound}) and comparing with (\ref{bound}). However,
as the value of $m_{\chi}$ is reduced, the range of masses for
which the tachyonic decay channel is open grows and begins to
dominate over that of narrow parametric resonance. Tachyonic decay
can be made to work even for $\mi \sim \mphi$.

For values of $\mphi$ for which both decay channels are open,
the tachyonic decay is more efficient for two reasons.
First, it leads
to the excitation of all long wavelength modes, and not just to
modes in a narrow resonance band. Modes with
\beq k_p^2 \, < \, \mi^2 {{\Phi} \over {M_p}} \, \equiv \,
k^2_{crit} \eeq
are excited. For such modes, the value of $\eta_k$ increases with
a maximal rate given by
\beq \label{growth2}
\eta_k \, \sim \, exp({\sqrt{q}} z) \, ,
\eeq
which for $q < 1$ is a larger rate than that which occurs for
modes in the resonance band during narrow resonance (see
(\ref{growth})). This is the second reason for the larger
efficiency of the tachyonic decay channel.

Let us estimate the energy density $\rho_{\eta}$ stored in the
quanta produced during the tachyonic decay process. The phase
space of modes which are excited tachyonically is of the order
$k_{crit}^3$. Each mode grows with a rate which varies in time,
the maximal rate being given by (\ref{growth2}), and the growth
occurs for approximately half the oscillation period (the period
during which the effective square mass is negative). The mean
growth rate is given by $1 / \sqrt{2}$ of the maximal rate. Thus
\beq
\rho_{\eta}(t) \, \sim \,
\mi^4 {{\Phi^2} \over {M_p}^2} exp({1 \over {\sqrt{8}}} \mi \sqrt{\Phi / M_P} t) \, .
\eeq
The prefactor in front of the exponential factor is much larger
than the corresponding factor in the case of the narrow resonance
decay process.

\section{Discussion and Conclusions}\label{discussions}

In this paper we have studied two non-perturbative decay processes
which can substantially dilute the density of dangerous moduli
fields. Both processes occur during the phase in which the moduli
fields oscillate about their ground states. The first
non-perturbative process is parametric resonant excitation of
fields coupled to the modulus field. In this paper we modeled such
fields as a scalar field $\chi$ coupled to the modulus field
$\phi$ via dimension five operators suppressed by the Planck mass.
The second decay process is tachyonic decay which makes use of the
fact that the effective square mass of $\chi$ in the presence of
the oscillating $\phi$ field is negative for part of the
oscillation period. These decay processes are analogous to the
parametric resonant decay of the inflaton during reheating
\cite{TB}, and to tachyonic preheating \cite{Felder},
respectively.

We have established the conditions under which either of the two
decay processes can solve the moduli problem, i.e. reduce the
energy density of the modulus field to values consistent with big
bang nucleosynthesis. It appears that narrow parametric resonance
has the potential to solve the modulus problem but, for the toy
model considered here, and using very conservative conditions
for the effectiveness of the resonance process, 
it happens only at the time when the pertubative
decay rate also becomes important. Undoubtedly, tachyonic decay
successfully solves the problem given masses of decay products 
which are
much smaller than the scale of interaction - $\mi$. Moreover, the
tachyonic decay channel allows for excitation of particles with
masses heavier than that of the decaying particle. Depending on the
values of the other masses in the Lagrangian, either of the two
decay processes can be open for a wider range of masses $\mphi$.
For values of $\mphi$ for which both decay channels are open, the
tachyonic decay is much more efficient, as is true in the case of
the decay of the inflaton. Note that the presence of an
interaction term in the Lagrangian linear in $\phi$ was important
in order to obtain the tachyonic decay.

It will be of great interest to study the applicability of these
decay channels to concrete models with moduli fields. This work is
left for future research.

The form of the potential in our toy model Lagrangian inevitably
suggests inflation at low scales. This natural source of inflation
does not only dilute heavier relics but could also mitigate the
flatness, horizon and entropy problems. It needs to be studied
whether this type of models can provide a successful reheating
mechanism, if the inflaton has only gravitationally suppressed
interactions. Once again, non-perturbative instabilities like
those used in preheating \cite{TB,KLS1,STB,KLS2,Felder} are likely
to be successful. If the modulus field comes to dominate the
energy density of the universe for some period (without
necessarily leading to inflation), it can provide a candidate for
the curvaton (see e.g. \cite{Anupam} for an extensive discussion
of moduli fields as candidates for the curvaton).

\acknowledgments

This work is supported by NSERC. The research of N.S. is also
supported in part by McGill University under a McGill Graduate
Student Fellowship.
We would like to thank the Perimeter Institute (N.S.)
and the Yukawa Institute for Theoretical Physics
at the University of Kyoto (and in particular Misao Sasaki) (R.B.) for
kind hospitality during the time when this project was completed.




\begin{thebibliography}{99}

\bibitem{Coughlan}
G.~D.~Coughlan, W.~Fischler, E.~W.~Kolb, S.~Raby and G.~G.~Ross,
Phys.\ Lett.\ B {\bf 131}, 59 (1983).

\bibitem{Ellis2}
J.~R.~Ellis, D.~V.~Nanopoulos and M.~Quiros,
Phys.\ Lett.\ B {\bf 174}, 176 (1986).

\bibitem{Carlos}
B.~de Carlos, J.~A.~Casas, F.~Quevedo and E.~Roulet,
Phys.\ Lett.\ B {\bf 318}, 447 (1993) [arXiv:hep-ph/9308325].

\bibitem{Banks:1993en}
T.~Banks, D.~B.~Kaplan and A.~E.~Nelson,
Phys.\ Rev.\ D {\bf 49}, 779 (1994) [arXiv:hep-ph/9308292].

\bibitem{Goncharov}
A.~S.~Goncharov, A.~D.~Linde and M.~I.~Vysotsky,
Phys.\ Lett.\ B {\bf 147}, 279 (1984).

\bibitem{Linde}
A.~D.~Linde,
Phys.\ Lett.\ B {\bf 116}, 335 (1982).

\bibitem{FordVil}
A.~Vilenkin and L.~H.~Ford,
Phys.\ Rev.\ D {\bf 26}, 1231 (1982).

\bibitem{Giudice2}
G.~F.~Giudice, A.~Riotto and I.~I.~Tkachev,
JHEP {\bf 0106}, 020 (2001) [arXiv:hep-ph/0103248].

\bibitem{Giudice1}
G.~F.~Giudice, I.~Tkachev and A.~Riotto,
JHEP {\bf 9908}, 009 (1999) [arXiv:hep-ph/9907510].

\bibitem{Felder2}
G.~N.~Felder, L.~Kofman and A.~D.~Linde,
JHEP {\bf 0002}, 027 (2000) [arXiv:hep-ph/9909508].

\bibitem{Ellis1}
J.~R.~Ellis, A.~D.~Linde and D.~V.~Nanopoulos,
Phys.\ Lett.\ B {\bf 118}, 59 (1982).

\bibitem{Dine:1995uk}
M.~Dine, L.~Randall and S.~Thomas,
Phys.\ Rev.\ Lett.\  {\bf 75}, 398 (1995) [arXiv:hep-ph/9503303].

\bibitem{Randall:1994fr}
L.~Randall and S.~Thomas,
Nucl.\ Phys.\ B {\bf 449}, 229 (1995) [arXiv:hep-ph/9407248].

\bibitem{Stewart}
D.~H.~Lyth and E.~D.~Stewart,
Phys.\ Rev.\ D {\bf 53}, 1784 (1996) [arXiv:hep-ph/9510204].

\bibitem{Asaka:1999xd}
T.~Asaka and M.~Kawasaki,
Phys.\ Rev.\ D {\bf 60}, 123509 (1999) [arXiv:hep-ph/9905467].

\bibitem{Kawasaki1}
T.~Asaka, M.~Kawasaki and M.~Yamaguchi,
Phys.\ Lett.\ B {\bf 451}, 317 (1999) [arXiv:hep-ph/9810334].

\bibitem{Kawasaki2} M.~Kawasaki and F.~Takahashi,
Phys.\ Lett.\ B {\bf 618}, 1 (2005) [arXiv:hep-ph/0410158].

\bibitem{Dine84}
M.~Dine, W.~Fischler and D.~Nemeschansky,
Phys.\ Lett.\ B {\bf 136}, 169 (1984).

\bibitem{Coughlan84}
G.~D.~Coughlan, R.~Holman, P.~Ramond and G.~G.~Ross,
Phys.\ Lett.\ B {\bf 140}, 44 (1984).

\bibitem{Linde96}
A.~D.~Linde,
Phys.\ Rev.\ D {\bf 53}, 4129 (1996) [arXiv:hep-th/9601083].

\bibitem{Kofman}
L.~Kofman, A.~Linde, X.~Liu, A.~Maloney, L.~McAllister and
E.~Silverstein,
JHEP {\bf 0405}, 030 (2004) [arXiv:hep-th/0403001].

\bibitem{Watson}
S.~Watson,
Phys.\ Rev.\ D {\bf 70}, 066005 (2004) [arXiv:hep-th/0404177].

\bibitem{TB}
J.~H.~Traschen and R.~H.~Brandenberger,
Phys.\ Rev.\ D {\bf 42}, 2491 (1990).

\bibitem{KLS1}
L.~Kofman, A.~D.~Linde and A.~A.~Starobinsky,
Phys.\ Rev.\ Lett.\  {\bf 73}, 3195 (1994) [arXiv:hep-th/9405187].

\bibitem{STB}
Y.~Shtanov, J.~H.~Traschen and R.~H.~Brandenberger,
Phys.\ Rev.\ D {\bf 51}, 5438 (1995) [arXiv:hep-ph/9407247].

\bibitem{KLS2}
L.~Kofman, A.~D.~Linde and A.~A.~Starobinsky,
Phys.\ Rev.\ D {\bf 56}, 3258 (1997) [arXiv:hep-ph/9704452].

\bibitem{Felder}
G.~N.~Felder, J.~Garcia-Bellido, P.~B.~Greene, L.~Kofman,
A.~D.~Linde and I.~Tkachev,
Phys.\ Rev.\ Lett.\  {\bf 87}, 011601 (2001)
[arXiv:hep-ph/0012142].

\bibitem{Fermionic}
P.~B.~Greene and L.~Kofman,
Phys.\ Lett.\ B {\bf 448}, 6 (1999) [arXiv:hep-ph/9807339].

\bibitem{Landau}
L. Landau and I. Lifshitz, {\it Mechanics} (Pergamon, Oxford, 1960).

\bibitem{Arnold}
V. Arnold, {\it Mathematical Methods of Classical Mechanics}
(Springer, New York, 1974).

\bibitem{Anupam}
K.~Enqvist and A.~Mazumdar,
  Phys.\ Rept.\  {\bf 380}, 99 (2003)
  [arXiv:hep-ph/0209244].

\end{thebibliography}
\end{document}